\documentclass[epj]{svjour}
\usepackage{graphics}
\usepackage{amsmath}
\usepackage{amsfonts}
\usepackage{xcolor}

\newcommand{\unc}[1]{\ensuremath{\mathcal{\hat{U}}(\mathcal{#1})}}

\begin{document}
\title{An iterative method to estimate the combinatorial background}
\subtitle{}
\author{Georgy Kornakov\inst{1}
\thanks{\emph{Present address:} G.Kornakov@gsi.de},%
Tetyana Galatyuk\inst{1,2}
}                     
\offprints{}          
\institute{Insert the first address here \and the second here}
\institute{$^{1}$Technische Universit\"at Darmstadt
	\\$^{2}$GSI Helmholtzzentrum f\"ur Schwerionenforschung}
%
\date{August 2018}
%
\abstract{
The reconstruction of broad resonances is important for understanding the dynamics of heavy ion collisions. However, large combinatorial background makes this objective very challenging. In this work an innovative iterative method which identifies signal and background contributions without input models for normalization constants is presented. This technique is successfully validated on a simulated thermal cocktail of resonances. This demonstrates that the iterative procedure is a powerful tool to reconstruct multi-differentially inclusive resonant signals in high multiplicity events as produced in heavy ion collisions.  
\PACS{
      {29.}{Experimental methods and instrumentation for elementary-particle and nuclear physics}   \and
      {29.85.-c}{Computer data analysis}
     } 
} 
\maketitle
\section{Introduction}
\label{intro}
In heavy ion collisions, one of the observables that is commonly used to characterize the properties of produced matter is the two-body invariant mass spectrum. The short-lived (unstable) particles appear as a bump with a width related to the lifetime of the particle, ($\Gamma\sim1/\tau$). In the case of excited baryon states like $\Delta$ , $N^*$ or the $\rho$ meson, their broad width and a poor signal-to-background ratio make the reconstruction difficult. The main challenge is the presence of a large combinatorial background due to random combinations of measured particles. 

The precise measurement of the differential inclusive spectra of short lived baryons and mesons would allow to improve the understanding of the mechanisms of particle production and to test available theoretical models by providing precise inclusive spectra. 

Several approaches have been proposed to solve this problem, among them can be named techniques like Monte-Carlo phase-space generators (integrators) \cite{phase_space_James:1968}, a common tool for inclusive analysis in elementary reactions. The like-sign method, which is commonly used for dilepton {\cite{likesign_1996,likesign_Marek_2000,likesign_SPS_2000,likesign_PHENIX_2010}} and dipion analysis at high energies \cite{likesign_STAR_rho0_Adams_2003}, can be proven to be exact for certain cases, but it cannot be applied universally. For many-particle events, methods based on position swapping have been proposed with good results for the diphoton invariant mass reconstruction \cite{position_swapping_vanEijndhoven:2001}. However, the most common technique is the Event-Mixing method \cite{event_mixing_uncorrelator_first_idea_Kopylov:1974th}, introduced originally to generate the reference distribution to study the identical particle correlations. The main idea is to combine particles from different collisions, generating an uncorrelated spectrum that after normalization is subtracted from the total same-event distribution, yielding the signal spectrum as their difference. Even though it has been pointed out that the background obtained with this technique has several problems due to the determination of normalization constants, the particle and event class selection or acceptance \cite{event_mixing_failing_Higgins:1978,event_mixing_LHote:1992,event_mixing_Oset_failing_Kaskulov:2010}, it is the most commonly applied method \cite{event_mixing_and_likesign_pippim_Jancso:1977,event_mixing_ISR_deltapp_Breakstone:1983,event_mixing_Drijard:1984,event_mixing_deltapp_Trzaska:1991,event_mixing_unfolding_eskef_FOPI:1998,event_mixing_high_mass_dileptons_Crochet:2001,event_mixing_like_sign_NA60_rho_Arnaldi:2006,event_mixing_STAR_phi_Abelev:2008}.

In general, decay products are used to generate the background as well as uncorrelated particles, having commonly different kinematics. Consequently, contributions from signal particles might distort the generated background. This is a major drawback, since the uncorrelated distribution will be usually different from the true combinatorial background \cite{event_mixing_LHote:1992}. One possible solution to overcome this is to generate the contribution from signal pairs to the combinatorial background by modelling the yield, spectral shape and decay kinematics of the studied resonance. However, it introduces a strong model dependence making the study of the modification of the line shapes impossible. In this paper, we describe a new unbiased method to estimate the correlated distribution directly from data. 

It is worthwhile to mention that a similar approach has been developed in gamma-spectroscopy in order to suppress contributions from Compton and backward scattering, where the sought signal is stripped or unfolded from the total measured gamma spectrum in a similar iterative procedure \cite{gamma_spectroscopy_Radford:1987}.

This paper is organized as follows: in Sec.~\ref{sec:method} the iterative method is presented. In Sec.~\ref{sec:analysis} a simulated cocktail of resonances and thermal sources is used to demonstrate the capabilities of the developed technique. In Sec.~\ref{sec:discussion} the method is discussed regarding its capabilities and limitations. Also, it is put into context of other techniques developed for background subtraction in other fields. Finally, the work is summarized.  

\section{The iterative method}

\label{sec:method}
Let $\mathcal{T}$ be the total discrete multi-differential spectrum obtained after adding all possible same-event four-momenta combinations of two particles of type 1 and 2 over all events.
	
The signal, $\mathcal{S}$, is comprised of those combinations of particles that are daughters of same mother, $R\rightarrow P_1 + P_2$. The remaining combinations belong to the combinatorial background $\mathcal{B}$ and they relate to each other as
\begin{equation}
\mathcal{T}=\mathcal{B}+\mathcal{S}.\label{eq_sig_eq_rsig_plus_bgd}
\end{equation}

The goal of the iterative method is to transform $\mathcal{T}$ such that only signal remains at the end. For that purpose, uncorrelating operators acting only on signal can be used. Then, let $\mathcal{\hat{U}}$ be a non-linear uncorrelating operator, for example, position swapping, event mixing or track rotation. The latter technique consists in modifying the orientation of single particle three-momentum preserving its module. It is widely used to calculate the combinatorial background beneath narrow peaks \cite{rotations_STAR_multistrange_Adams:2005,rotations_STAR_antimatter_Abelev:2010,rotations_CERES_lambda_Adamova:2012,rotations_ALICE_JPsi_Abelev:2012} and is the operator of choice in this work.   

\begin{figure}
	\centering
	\resizebox{0.5\textwidth}{!}{%
	\includegraphics{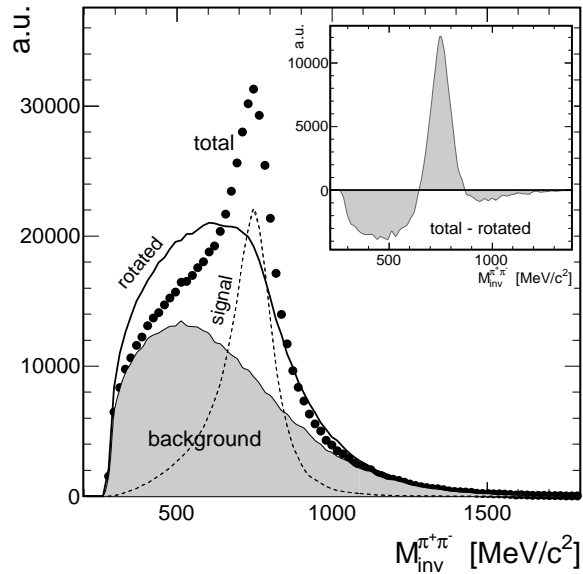}
    }
	\caption{Example of an invariant mass distribution (full circles), $\mathcal{T}$, which is comprised of signal $\mathcal{S}$ (dashed line) and combinatorial background $\mathcal{B}$ (grey area). The solid line is the uncorrelated $\unc{T}$ distribution obtained with the rotation technique (see for details in text). The inset shows the invariant mass of the $\mathcal{T}-\unc{T}$ distribution. Notice the positive and negative areas coinciding with the peak location and its sides.}
	\label{fig_definitions_of_bgd_sig_trans}
\end{figure}

In order to illustrate these definitions a simulated set of three uncorrelated $\rho^{0}(770)$ per event decaying into pion pairs has been produced. The proportion of signal pairs to random combinations is 3:6. The invariant mass projections for total, signal, background and the total rotated spectra are shown in  Fig.~\ref{fig_definitions_of_bgd_sig_trans}. The invariant mass projections of the difference between $\mathcal{T}$ and $\unc{T}$ is shown in the inset of the same figure. As pairs from background are insensitive to rotations, these differences are only due to signal. After rotating one track from the pair, the signal spreads over a broader phase space region. Hence, positive values are expected in the peak and negative values on the sides. Provided a model exists for the signal, it is possible to extract directly the correct signal yield, whose $\mathcal{S}-\unc{S}$, saturates the $\mathcal{T}-\unc{T}$ distribution. Although the reconstruction would be biased \cite{event_mixing_LHote:1992}.  

Alternatively, the signal can be approached by an iterative solution. The positive difference areas of $\mathcal{T}-\unc{T}$ are due to the signal. The first iterative solution, $\mathcal{S^{\text{0}}}$, corresponds to these regions. Then, the uncorrelator operator is applied upon the first solution and the improved approximate solution $\mathcal{S^{\text{1}}} = \mathcal{T}-(\unc{T}-\unc{S^{\text{0}}})$ is obtained. This process repeats until complete saturation of the total difference distribution. 

Mathematically, the proposed algorithm can be considered to be similar to the Landweber Iterative Method \cite{Landweber:1951}, widely used to solve ill-conditioned, noisy and non-linear systems \cite{landweber_method_SCHERZER:1995,landweber_discrete_QinianJin:2001}. From Eq.~\ref{eq_sig_eq_rsig_plus_bgd}, $\mathcal{B}=\unc{B}$ and the fact that the positive areas of $\mathcal{T} - \unc{T}$ belong to signal, the iterative solution for the $\mathcal{S}$ matrix can be written as 
\begin{equation}
\mathcal{S}^{k+1} = \text{max}\{ \mathcal{T} - (\unc{T} - \unc{S^{\mathit{k}}}),0\} ,\label{formula_iteration}
\end{equation}
where only positive values are chosen and \textit{k} is the iteration index. Then, convergence is reached once 
\begin{equation}
\mathcal{S}^{k+1} = \mathcal{S}^k.
\end{equation}
This is exact for infinite statistics. Otherwise, the bin-to-bin statistical fluctuations introduce a systematic increment of the signal with every iteration. This increment depends on the number of pairs contributing to the total matrix $\mathcal{T}$ and those to $\unc{B}$. As a consequence, $\mathcal{S}^{k}$ will grow continuously and eventually will reach $\mathcal{T}$. A simple example can be a data set without signal in it. In such a case, after applying the uncorrelating operator and subtracting the produced distribution from the total approximately half of the bins will have a positive value. This phony signal will increment further with each iteration the entries in $\mathcal{S}^{k}$. The stopping criteria for the iterator in case of finite statistics is given by the change in the signal increment rate as it moves from the real signal domain towards the statistical increment regime. There is no common stop iteration for all the bins together. Individual values have to be extracted by identifying the transition from increase due to real signal to increase due to statistical fluctuations. In the abovementioned example without signal, the increment rate will be constant and only depending on the number of entries in each bin.

\section{Method application to simulated data}
\label{sec:analysis}

\begin{figure}
	\centering
	\resizebox{0.5\textwidth}{!}{%
		\includegraphics{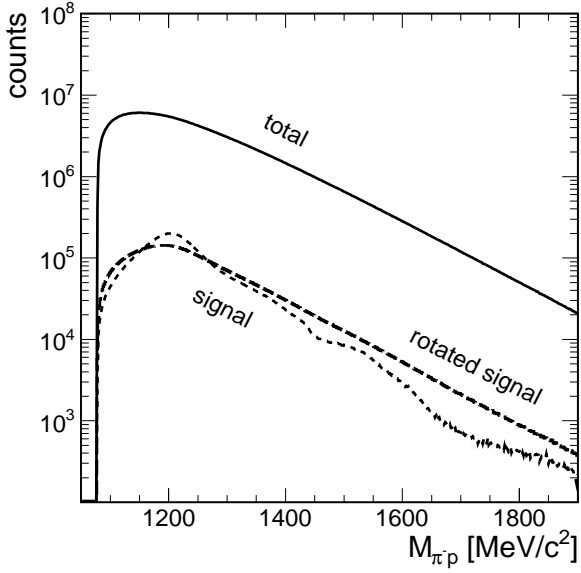}
	}
	\caption{Signal generated with Pluto \cite{pluto_Frohlich:2007} imitating the environment of a heavy ion collision. The solid curve is the total invariant mass distribution $\mathcal{T}$, the short-dashed line is the resonant signal $\mathcal{S}$, and the long-dashed line is the uncorrelated resonant signal distribution $\unc{S}$. The signal-to-background ratio is below 5~\%. The $\unc{T}$ invariant mass distribution is not shown as it would be indistinguishable in this representation and equal to $\mathcal{T}$.}
	\label{All_and_sig_and_sig_rot}
\end{figure}

In order to present the complete picture, a simulation consisting of pions and protons from a thermal source together with a cocktail of resonances decaying into a proton and a negative pion in the final state was produced with Pluto \cite{pluto_Frohlich:2007}, mimicking particle production in a heavy ion collision at 1.25 GeV per nucleon. The resonances were generated assuming a thermal source of temperature of 65 MeV, their relative abundance, decay channels and branching ratios are summarized in Table~\ref{table_PLUTO_cocktail}. The invariant mass distribution of $\mathcal{T}$, $\mathcal{S}$ and $\unc{S}$ are shown in Fig.~\ref{All_and_sig_and_sig_rot}, having a signal-to-background ratio below 5\%.

\begin{table}
	\caption{Cocktail of particles generated by Pluto \cite{pluto_Frohlich:2007} to imitate the environment of a heavy ion collision at intermediate energies. A thermal source of 65 MeV was considered.}
	\label{table_PLUTO_cocktail}
	\begin{tabular}{c|ccc}
		\hline\noalign{\smallskip}
		& N/Nprot & Channels & $\Gamma/\Gamma_i$ \\ 
		\noalign{\smallskip}\hline\noalign{\smallskip}
		$\Delta(1232)^0$ & 5 \%  & $\pi^-$p & 1 \\ 
		$N(1440)^0$ & 2 \%  & $\pi^-$p, p$\rho^{-}$,$\Delta^+\pi^-$ & 0.62/0.02/0.36 \\ 
		$N(1535)^0$ & 1.5 \% & $\pi^-$p, p$\rho^{-}$,$\Delta^+\pi^-$ & 0.70/0.06/0.23 \\ 
		$\Delta(1620)^0$ & 1 \%  & $\pi^-$p, p$\rho^{-}$,$\Delta^+\pi^-$ & 0.18/0.12/0.70\\ 
		$\Delta(1920)^0$ & 0.5 \%  & $\pi^-$p, p$\rho^{-}$,$\Delta^+\pi^-$ & 0.7/0.2/0.1 \\ 
		p & 20 &  &  \\ 
		$\pi^-$ & 10 \%  &  & \\
		\noalign{\smallskip}\hline
	\end{tabular}
\end{table}

\begin{figure}
	\centering
	\resizebox{0.5\textwidth}{!}{%
		\includegraphics{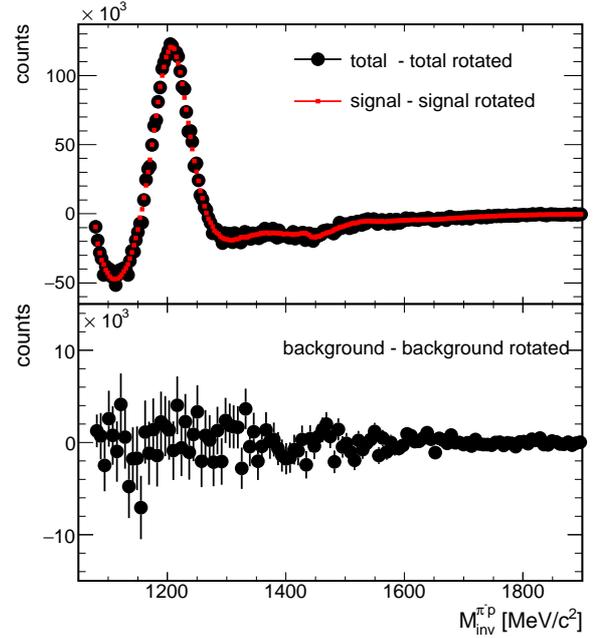}
	}
	\caption{Top panel, invariant mass distribution $\mathcal{T}-\unc{T}$ (black circles) and $\mathcal{S}-\unc{S}$ distribution (red dots) obtained from the cocktail simulation. The non-zero difference is only due to the signal contribution. Bottom panel, invariant mass distribution of $\mathcal{B}-\unc{B}$, which is compatible with zero within statistical uncertainties.}
	\label{fig_differences_sig_sigr_rsig_rsigr}
\end{figure}

\begin{figure}
	\centering
	\resizebox{0.5\textwidth}{!}{%
		\includegraphics{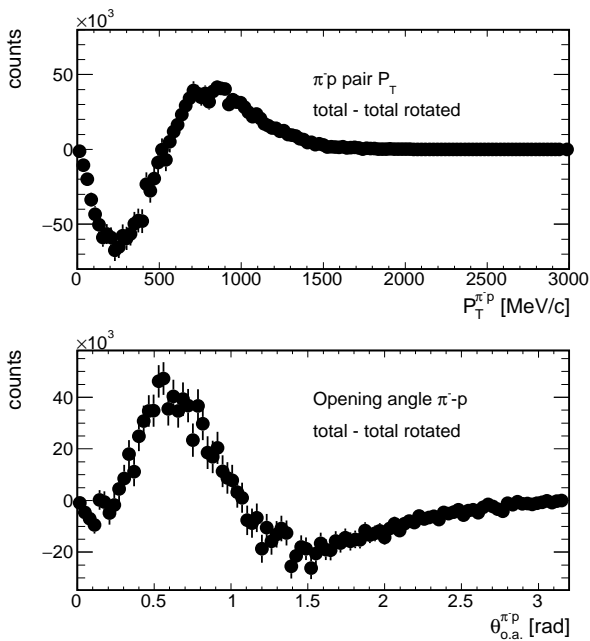}
	}
	\caption{Projections of $\mathcal{T}-\unc{T}$ from cocktail simulation to the transverse momentum (top) and opening angle (bottom). These projections point towards the necessity of multi-differential analysis in order to account for differences in the signal and background in different phase-space bins.}
	\label{fig_differences_opangle_pt}
\end{figure}

The $\mathcal{T}-\unc{T}$, $\mathcal{S}-\unc{S}$ and $\mathcal{B}-\unc{B}$ difference  distributions of the invariant mass of $\pi^-p$ pairs are shown in Fig.~\ref{fig_differences_sig_sigr_rsig_rsigr}. It is evident from this figure that the difference in the shape is only due to the signal and that the background does not contribute to the total difference. Fig.~\ref{fig_differences_opangle_pt} shows the difference distributions of the opening angle $\alpha$ and pair transverse momentum $P_{T}$. The non-uniformity points to the necessity of a differential analysis. In previous works \cite{event_mixing_Oset_failing_Kaskulov:2010}, it has already been observed that different particle or phase-space selections were modifying the shape of the reconstructed background and it has to be accounted for during the reconstruction. Therefore, the reconstruction of the signal is performed in four dimensions: invariant mass, pair transverse momentum, pair rapidity and opening angle. The isotropy of the azimuthal emission angle and the decay plane are assumed. Otherwise, these extra two dimensions have to be included as well in the reconstruction.

\begin{figure}
	\resizebox{0.5\textwidth}{!}{%
		\includegraphics{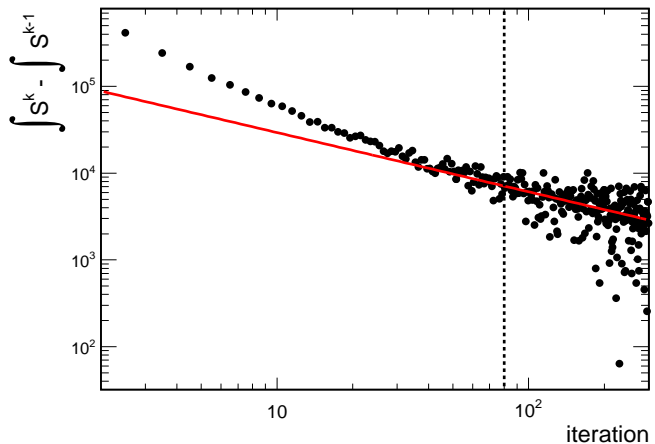}
	}
	\caption{Integrated increment of the signal yield in every iteration step. The change in slope indicates a transition between the regime with real signal growth and the rise of signal due to bin-to-bin statistical fluctuations. In this example the transition happens between the 40th and the 80th iteration.}
	\label{fig_signal_increment}
\end{figure}

The starting minimization hypothesis is absence of signal, $\mathcal{S}=0$. The available phase-space is binned as follows: 40 in $M_{\pi^{-}p}$, 5 in $P_T$, 5 in $\alpha$ and 5 in $Y$. Then, following the Eq.~\ref{formula_iteration}, the first solution $\mathcal{S}^0$ is given by $\text{max}\{\mathcal{T}-\unc{T},0\}$. The next iteration includes the contribution of the signal obtained in the first step after rotating the pairs from it. For the simulated data sample 300 iterations have been performed. The increment in the total yield of the signal is shown in Figure~\ref{fig_signal_increment}. The largest increase happens in the first iterations as expected. In this particular case, an ankle around the 80th iteration indicates that the statistical increment regime was reached and no more resonant particle pairs can be incorporated into the reconstructed signal.

\begin{figure}
	\resizebox{0.5\textwidth}{!}{%
		\includegraphics{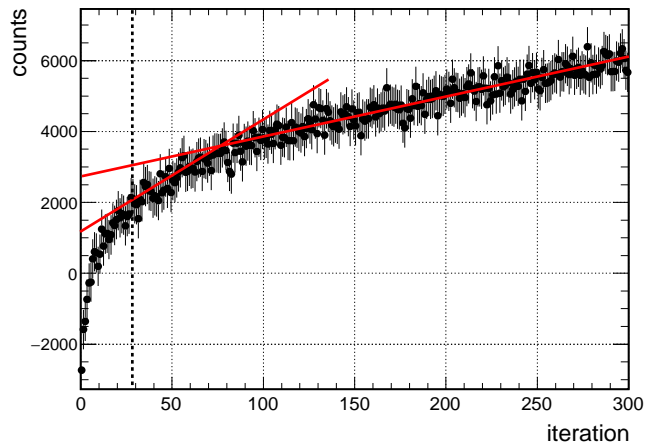}
	}
	\caption{Signal yield evolution for each iteration step for a given $M$ and $P_T$ bin, integrated over the whole rapidity and pair opening angle. The true signal, in this phase-space bin, grows only until the 28th iteration whereas is incompatible with a linear trend.}
	\label{fig_evolution_one_bin_s_val}
\end{figure}

However, as it was introduced in Sec.~\ref{sec:method}, it is not enough to stop the iterator at this value. Due to different signal to background ratios over the available phase-space, the yield evolution has to be studied independently for each bin. Three regions have to be identified, the first with a fast rising, then an increase that linearly continues until the ankle found in the total increment evolution and finally the continuous increase, as shown in Fig.~\ref{fig_evolution_one_bin_s_val}. An algorithm finds the iteration at which the increase becomes compatible with linear in the second region. After this procedure is applied for all bins, one more iteration should be performed to ensure that the condition $\mathcal{T}-\unc{T} = \mathcal{S}-\unc{S}$ is fulfilled. 

\begin{figure}
	\resizebox{0.5\textwidth}{!}{%
		\includegraphics{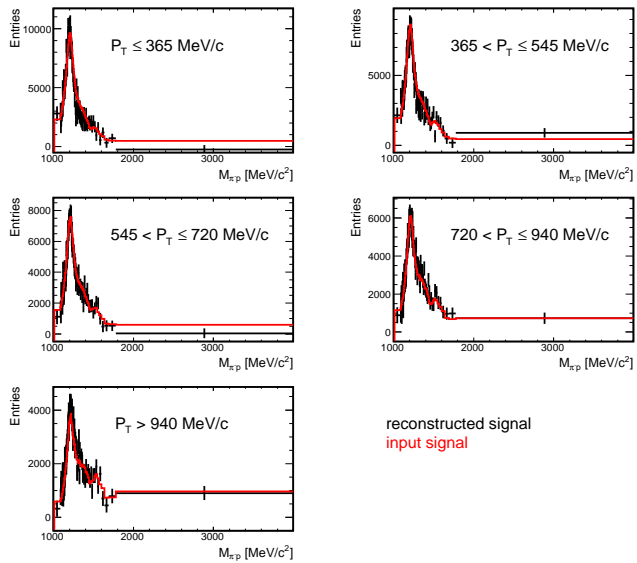}
	}
	\caption{Differential $\pi^-p$ invariant mass distribution in 5 $P_T$ bins after bin-by-bin convergence. The input resonance signal from Pluto is shown in red.}
	\label{fig_result_differential_pt}
\end{figure}

\begin{figure}[!htbp]
	\resizebox{0.5\textwidth}{!}{%
		\includegraphics{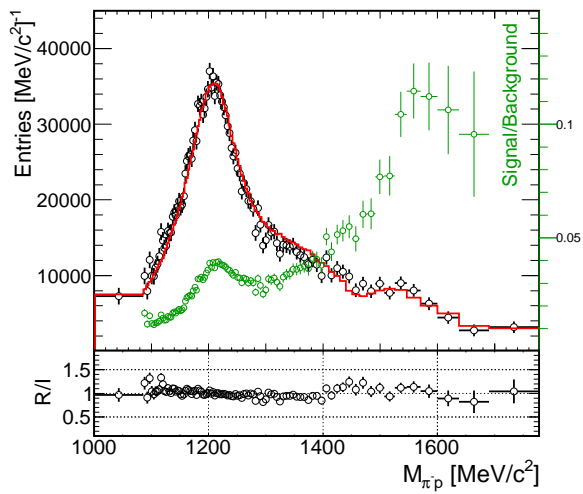}
	}
	\caption{Final integrated reconstructed invariant mass distribution (open circles) after applying the iterative procedure compared to the input distribution (red line). The signal to background ratio (green open circles). The ratio between input and reconstructed signals (R/I), shows agreement within statistical uncertainties, which are of the order of 10\%.}
	\label{fig_result_signal}
\end{figure}

Finally the differential and integrated solutions are obtained. In this example, with a signal-to-background ratio smaller than 5\%, the reconstructed and the input differential invariant mass are shown in Fig.~\ref{fig_result_differential_pt}. The integrated signal can be resolved with an accuracy better than 10\%, within statistical uncertainties, as shown in Fig.~\ref{fig_result_signal}. 

\section{Discussion and summary}
\label{sec:discussion}
The inclusive reconstruction of short lived states {($\tau\approx10^{-23}$~s)} is technically challenging due to the existence of a large combinatorial background. In this work the development of a model-independent and normalization-free iterative method has been addressed, in order to overcome the difficulties that other techniques have \cite{event_mixing_LHote:1992}. This method will facilitate a precise multi-differential identification of signal and background contributions to the reconstructed spectra.

The difference of the total and the generated uncorrelated distributions from the same data sample preserve information about the resonant signal. It was found that the positive areas belong to the signal distribution. The solution is obtained in an iterative approach, similar to the Landweber Iterative Method used to solve ill-conditioned, noisy and non-linear systems \cite{Landweber:1951,landweber_method_SCHERZER:1995}.

As it happened with the Landweber Iterative Method, the first solutions were slow and inefficient. There are many possibilities to be explored in order to improve the minimization: improved convergence criteria to stop earlier the number of iterations, the pre-conditioning of the initial matrices, and the accelerated or boosted versions with larger gradients.  

Most of the effects of a real measurement have not been considered such as finite resolution, wrong particle identification, efficiency, acceptance nor occupancy. A non-uniform probability distribution of the rotation angle can account for them. The needed corrections can be derived from data.

The algorithm has been validated successfully on a simulated cocktail of thermal resonances. The mu\-lti-di\-ffe\-ren\-tial signal has been reconstructed, with a signal-to-background ratio below 5\%, in all mass ranges with an accuracy better that 10\% within statistical uncertainties.

The method capability is limited by the statistical fluctuations of the binned background. The more the broad signal approaches in shape the combinatorial background the larger becomes the number of needed iterations. This demands to perform a feasibility pre-study of the different distributions before applying the technique. Even continuum signals like $c\bar{c}$ or QGP signaltures like $q\bar{q}\rightarrow e^+e^-$ should be accessible.

The application of this technique would allow to reconstruct differentially inclusive spectra in both elementary and heavy ion collisions for the study of resonance production and their contribution to the yields  and differential spectra of stable particles measured by detectors. The reconstruction of $\pi^{\pm} p$ and $\pi^+\pi^-$ channels measured in Au+Au collisions at $\sqrt{s}=2.42$ GeV in HADES is ongoing. 

The authors wish to acknowledge stimulating discussions with M. Gumberidze, S. Harabasz, R. Holzmann, C. M\"untz, V. Pechenov, O. Pechenova, B. Ramstein, P. Salabura, H. Str\"obele and J. Stroth. This work has been supported by VH-NG-823 and Helmholtz Alliance HA216\-EMMI. 

%

\end{document}